# Dynamic optimization of volatile fatty acids to enrich biohydrogen production using a deep learning neural network

Mahmood Mahmoodi-Eshkaftaki [a*], Gustavo Mockaitis [b], Mohammad Rafie Rafiee [c]

[a] Department of Mechanical Engineering of Biosystems, Jahrom University, P.O. Box 74135-111, Jahrom, Iran.

[b] Interdisciplinary Research Group on Biotechnology Applied to the Agriculture and the Environment, School of Agricultural Engineering, University of Campinas (GBMA/FEAGRI/UNICAMP), 501 Candido Rondon Avenue, CEP, 13.083-875, Campinas, SP, Brazil.

[c] Department of Water Sciences & Engineering, Jahrom University, P.O. Box 74137-66171, Jahrom, Iran.

**\* Email:** m.mahmoodi5@gmail.com, m.mahmoodi5@jahromu.ac.ir

**Abstract**

A new strategy was developed to investigate the effect of volatile fatty acids (VFAs) on the efficiency of biogas production with a focus on improving bio-$H_2$. The inoculum used, anaerobic granular sludge obtained from a UASB reactor treating poultry slaughterhouse wastewater, was pretreated with four different pretreatments. The relationship between VFAs and biogas compounds was studied as time-dependent components. In time-dependent processes with small sample size data, regression models may not be good enough at estimating responses. Therefore, a deep learning neural network (DNN) model was developed to estimate the biogas compounds based on the VFAs. Accuracy of this model to predict the biogas compounds was higher than that of multivariate regression models. Further, it could predict the effect of time changes on biogas compounds. Analysis showed that all the pretreatments were able to increase the ratio of butyric acid / acetic acid successfully, decrease propionic acid drastically, and increase the efficiency of bio-$H_2$ production. As discovered, butyric acid had the greatest effect on bio-$H_2$, and propionic acid had the greatest effect on $CH_4$ production. The best amounts of the VFAs were determined using an optimization method, integrated DNN and desirability analysis, dynamically retrained based on digestion time. Accordingly, optimal ranges of acetic, propionic, and butyric acids were 823.2–1534.3, 36.3–47.4, and 1522–1822 mg/L, respectively, determined for digestion time of 25.23–123.63 h. These values resulted in the production of bio-$H_2$, $N_2$, $CO_2$, and $CH_4$ in ranges of 6.4–26.2, 12.2–43.2, 5–25.3, and 0–1.4 mmol/L, respectively. The optimum ranges of VFAs are relatively wide ranges and practically can be used in biogas plants.

# 1. Introduction

The production of waste materials is an undeniable part of human society. These wastes are produced by almost all anthropogenic activities including industries, forestry, agriculture, and municipalities. Anaerobic digestion (AD) is a suitable technology to convert wastes into valuable resources, produce biogas as a clean energy source, and reduce waste volumes and environmental impacts (Taherzadeh and Karimi, 2008). Agricultural by-products are the most important wastes in terms of production volume, and a rich source of raw materials for bioprocessing. Most of the agricultural wastes are comprised of cellulose, hemicellulose, and lignin, and must pass through a physicochemical pretreatment as upstream processing to hydrolysate and separate these compounds into fermentable monomers (Paudel et al., 2017). During these upstream pretreatment processes, the cellulose (35–45%) can be hydrolyzed into monosaccharides (mainly glucose) and separated from the hemicellulose (25–40%). Hemicellulose is a heteropolymer of hexose and pentose sugars, with D-xylose as the major constituent. Xylose is the second most abundant carbohydrate monomer after glucose in agricultural wastes and can be used to produce second-generation biofuels (Mockaitis et al., 2020). Whereas glucose can be efficiently converted to ethanol and fine chemicals, economical application for xylose needs development.

Bio-$H_2$ production from biomass wastes via dark fermentation not only addresses the energy demand in a renewable manner but also resolves the safe disposal issues associated with these bio-wastes (Rambabu et al., 2020). Bio-$H_2$ is a clean product that can be used as an energy source to fuel combustion engines and fuel cells to produce energy. Energy obtained from bio-$H_2$ does not release carbon dioxide, and hydrogen has the highest energetic content (122 kJ/g) as compared with hydrocarbon fuels (Mahmoodi-Eshkaftaki and Mockaitis, 2022). Thus, the feasibility of the production of bio-$H_2$ through dark fermentation will be attainable only by optimizing the key parameters influencing the production process. Although the parameters affecting biogas plants are time-dependent (Radjaram and Saravanane, 2011; Thanwised et al. 2012; Lin et al., 2012; Kegl and Kralj, 2020), these parameters have been modeled as time-independent in the literature (Mahmoodi-Eshkaftaki and Rahmanian-Koushkaki, 2020; Mahmoodi-Eshkaftaki and Ebrahimi, 2021).

Albeit some studies have analyzed the effects of crucial physicochemical factors such as temperature, pH, and substrate concentration on bio-$H_2$ generation, few papers discuss the role of volatile fatty acids (VFAs) on bio-$H_2$ production. The VFAs, obtained as by-products from the AD, can be transformed into liquid biofuels such as ethanol and butanol or converted into biogas compounds such as $H_2$, $CH_4$, $CO_2$, and $H_2S$ (Steinbusch et al., 2008). The main VFAs typically found in the digestate are acetic, propionic n-valeric and butyric acids. Izumi et al. (2010) found that particle size reduction (for example using pretreatment process) can accelerate the production of soluble organic materials such as VFAs, increasing the possibilities of imbalance in the production and consumption of VFAs. Despite dark fermentation is a high-potential technology to process wastes into bio-$H_2$ (Mockaitis et al., 2020), it needs to be improved with the optimum fermentation condition.

Optimization based on multivariate regression models has been used to determine the optimal conditions of AD, (Mahmoodi-Eshkaftaki and Ebrahimi, 2019; Mahmoodi-Eshkaftaki and Rahmanian-Koushkaki, 2020). These methods need a high sample size data or low dimensionality. For processes with a small sample dataset or high dimensionality, especially with time-dependent variables, the regression models are not able to provide high accuracy for response estimation, so they cannot be used in the optimization (Ranjan et al., 2011). Supervised ANN models usually produce high accuracy but need big sample size data (Fan et al., 2017). Due to the high cost of properties measurement experiments in biological processes, the input dataset usually has small sample size data. Moreover, as the AD parameters are time-dependent, the simple ANN models cannot be used well. Therefore, using a deep learning neural network (DNN) instead of ANN and regression models can improve the optimization process (Ward et al., 2016; Mahmoodi-Eshkaftaki and Ebrahimi, 2021). Due to the difficulties of training DNN and assembling big datasets in material science, the use of DNNs in material science is limited. In recent years, the development of machine learning has provided some useful tools for the improvement of DNN, such as the use of genetic algorithms (GA) to train network weights and biases (Goodfellow et al., 2016) to deal with small datasets.

To optimize multiple responses, integrated accurate regression models and desirability analysis have been successfully used in previous studies (Wang et al., 2019; Mahmoodi-Eshkaftaki and Rahmanian-Koushkaki, 2020). However, for processes where the accuracy of regression models is low, integrated accurate neural network models and desirability analysis can offer reliable optimum amounts. Therefore, the present study aims to develop a DNN model according to the VFAs (input factors) to estimate $H_2$, $CH_4$, $N_2$, and $CO_2$ production (responses) in a time-dependent space and determine optimum amounts of the VFAs and corresponding biogas compounds using a dynamic multi-objective optimization method.

## 2. Materials and methods

### 2.1. Inoculum preparation

The seed inoculum used in this study was an anaerobic granular sludge obtained from a UASB reactor treating poultry slaughterhouse wastewater. The total volatile solids concentration (T.VS) in the sludge was 32 g/L, with a pH of 7.1. In the experiments, the sludge granules were macerated to improve the external mass transfer, increase superficial area of the bioparticles, and make the pretreatments more effective. Four different pretreatments were performed on the inoculum including acidic, thermal, acidic-thermal, and thermal-acidic compared with a control experiment. Inoculum pretreatments were carried out to select hydrogen producers from anaerobic sludge. Acidic pretreatment was performed by adding HCl solution (1 mol/L) until the pH of the sludge dropped down to 3.0 and remained unchanged under these conditions for 24 h. Afterwards, the pH was raised to 6.0 with the addition of NaOH solution (1 mol/L). The thermal pretreatment was done by raising the temperature of the sludge up to 90 °C for 20 min, and then the sludge was quickly cooled down to 25 °C with a water bath. Acidic-thermal and thermal-acidic pretreatments were a combination of acid and thermal pretreatments described above, carried out in sequence, after 24 h at room temperature between the two pretreatments (Mockaitis et al., 2019; Mockaitis et al., 2020). The control experiments were done without performing any pretreatment, directly using the macerated seed inoculum. For all pretreated and control sludges, the pH was set to 6.5 before their inoculation in the experiments. The pretreated sludge was kept at room temperature for 24 h before inoculation.

## *2.2. Experimental setup*

The medium used in the experiments contained xylose as the sole carbon source in all assays (7.5 ± 0.6 g/L). Nutrient supplementation was prepared as described in Mockaitis et al. (2020). All the experiments were performed in 500 mL Duran™ Flasks (total volume of 620 ± 14 mL), with a working volume of 285 ± 8 mL (medium + inoculum), and a headspace volume of 335 ± 18 mL. The headspaces of the reactors were purged for 3 min with $N_2$ 99% and sealed immediately (Wang et al., 2012). In all the assays, the initial inoculum concentration was 6.6 ± 0.3 g T.VS/L, and the initial pH was 6.68. The experiments were carried out under mesophilic conditions (30 °C) in a shaker incubator. Each assay lasted 169 h, and sampling was performed in 35-time steps. The VFAs of acetic, propionic, and butyric acids were measured for the assays, which influenced the response parameters. The responses were the biogas compounds of $H_2$, $N_2$, $CH_4$, and $CO_2$. The summary of the assays is reported in Table 1.

Table 1 – Statistical information of the experiments including digestion time, VFAs (time-dependent variables), and biogas compounds (time- and VFAs- dependent variables).

| Factors | Parameters | Range | Normalized | Mean | Std. Dev. |
|---|---|---|---|---|---|
| $t$ | Digestion time (h) | 0–168.2 | - | 56.096 | 40.381 |
| $x_1(t)$ | Acetic acid (mg/L) | 3–1672 | $-1 \sim 1$ | 551.993 | 401.259 |
| $x_2(t)$ | Propionic acid (mg/L) | 0–282 | $-1 \sim 1$ | 25.225 | 55.179 |
| $x_3(t)$ | Butyric acid (mg/L) | 3–1984 | $-1 \sim 1$ | 806.249 | 562.714 |
| $y_1(x,t)$ | $H_2$ (mmol/L) | 0–36.9 | $-1 \sim 1$ | 10.581 | 9.243 |
| $y_2(x,t)$ | $N_2$ (mmol/L) | 0–17.1 | $-1 \sim 1$ | 1.686 | 3.936 |
| $y_3(x,t)$ | $CO_2$ (mmol/L) | 0–35.8 | $-1 \sim 1$ | 17.349 | 9.266 |
| $y_4(x,t)$ | $CH_4$ (mmol/L) | 0–47.5 | $-1 \sim 1$ | 17.224 | 12.174 |

*2.3. DNN modeling*

The input dataset for DNN modeling contained three training parameters (including the VFAs) and four target parameters (the biogas compounds), which were repeated five times (control and pretreatment assays). The VFAs and biogas compounds were determined for 35 consecutive times from 0–168.2 h. Therefore, a DNN model was designed to predict biogas compounds based on VFAs, as the input and target parameters were time-dependent. DNN model design was performed using the software MATLAB (Ver. 7.8). The initial structure of the DNN consisted of an input layer with three neurons, an output layer with one neuron, and a hidden layer with 16 neurons. The details of the processes are presented in the flowchart and schematic diagram in Fig. 1.

To improve generalization and reduce the possibility of over-fitting in our neural network model, mean squared error (MSE) was used in the GA as the fitness function. The weights and biases of DNN layers were trained by integrating GA and neural network using the VFAs dataset of the first-time step (1×3), repeated for all assays (5). This fine-tuning DNN involved the minimization of MSE through regulating weights and biases. To train the DNN with $n$ hidden neurons using GA, $(number\ of\ input\ parameters) \times n + n + n + 1$ quantities are required in the weights and biases column vector. Therefore, in this study, for the input weights (3×n), the input biases (n), the output weights (n), and the output bias (1), 16 quantities were needed to design the GA dimension. The network optimization procedure by the GA was completed when the average relative changes in the best fitness function value over 100 generations was $\leq 1e-10$, the fitness function attained the value of $1e-10$, or the generation increased over 150. The effect of time on the parameters was trained using the trained recurrent neural network and updating the network state using the next time step. This process was repeated until the latest time step was achieved. Similar methods were used in previous studies to improve the ANN accuracy for the estimation of other parameters (Liu et al., 2017; Feng et al., 2019). They reported that these training methods of neural networks were suitable for the prediction of high dimensional low sample size datasets.

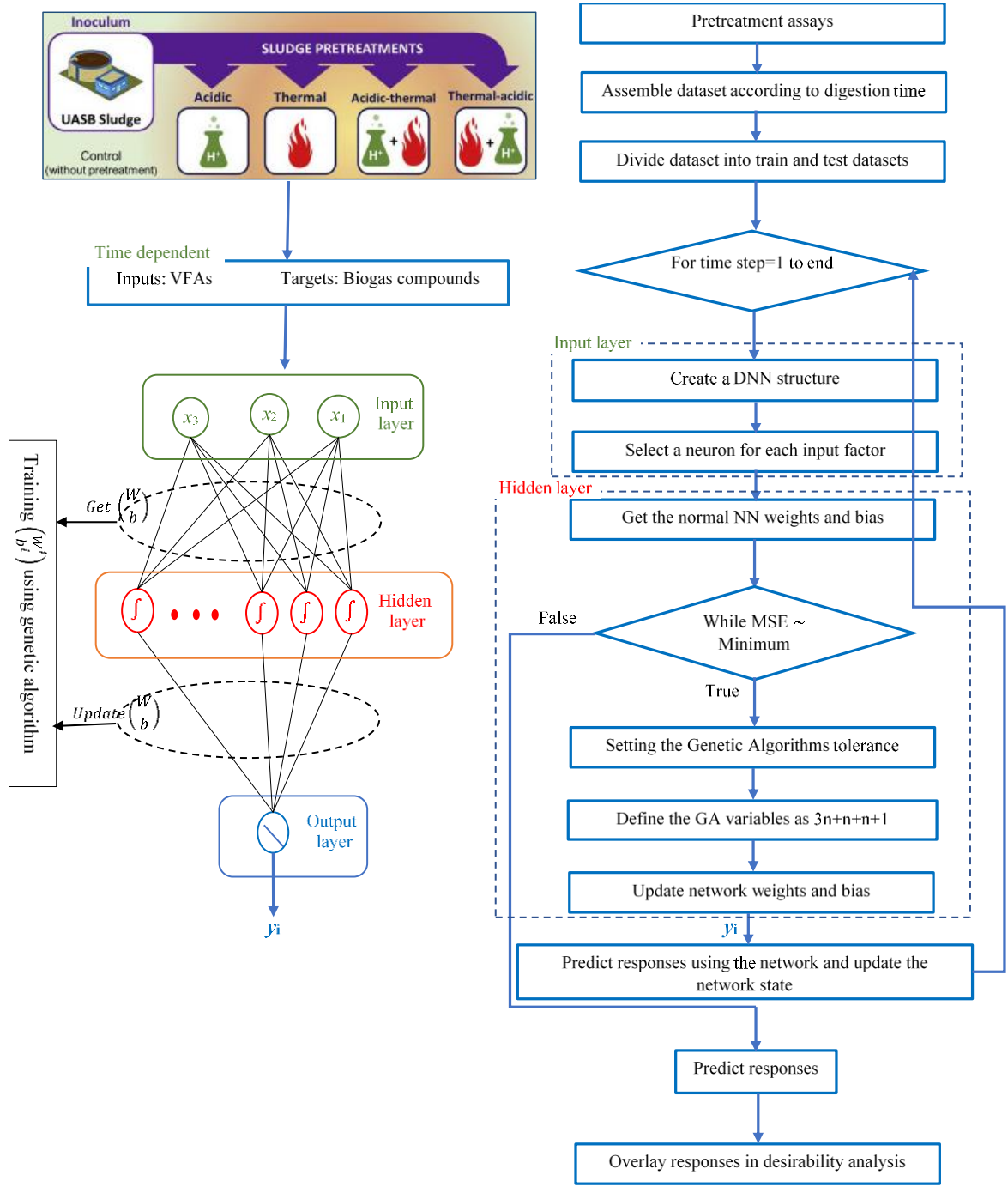

Figure 1 – Schematic diagram including the processes of pretreatment, data developing, DNN initializing and training, DNN retraining according to time, predict responses, and overlay the responses in desirability function.

## 2.4. Optimization process

This study introduces an efficient optimization method by combining DNN and desirability function to find the optimum amounts of VFAs and produce high amounts of bio-$H_2$ and $CH_4$, and low amounts of $CO_2$ and $N_2$. The responses, predicted with the DNN model, were overlayed in the desirability function to determine the optimum amounts. The desirability function was used to simultaneously optimize all affecting parameters to achieve the best conditions for AD. The method finds operating conditions $x_i$ providing the most desirable response values (Wang et al., 2019). For each response $y_i(x)$, the desirability function $d_i(y_i)$ was calculated using Eq.1. $d_i(y_i) = 0$ represented an undesirable value of $y_i$, and $d_i(y_i) = 1$ represented a highly desirable value. The overall desirability (D) was calculated using the geometric mean of the individual desirability levels. Predicted response values by DNN ($\hat{y}_i$) were used in place of the $y_i$ as suggested in the literature (Mahmoodi-Eshkaftaki and Ebrahimi, 2021).

$$D = \left[\prod_{i=1}^{n} d_i(\hat{y}_i)^{r_i}\right]^{\frac{1}{\sum r_i}}$$

$$d_i(\hat{y}_i) = \begin{cases} 0 & \text{if } \hat{y}_i(x) < L_i \\ \left(\frac{\hat{y}_i(x) - L_i}{T_i - L_i}\right) & \text{if } L_i \leq \hat{y}_i(x) \leq T_i \\ 1 & \text{if } \hat{y}_i(x) > T_i \end{cases}$$

If a response is to be maximized:
$T_i$ denoting a large enough value for the response.
$T_i = average\ (\hat{y}_i) + standard\ deviation\ (\hat{y}_i);$
$L_i = average\ (\hat{y}_i) - standard\ deviation\ (\hat{y}_i);$

$$d_i(\hat{y}_i) = \begin{cases} 1 & \text{if } \hat{y}_i(x) < T_i \\ \left(\frac{\hat{y}_i(x) - U_i}{T_i - U_i}\right) & \text{if } T_i \leq \hat{y}_i(x) \leq U_i \\ 0 & \text{if } \hat{y}_i(x) > U_i \end{cases}$$

If a response is to be minimized:
$T_i$ denoting a small enough value for the response.
$U_i = average\ (\hat{y}_i) + standard\ deviation\ (\hat{y}_i);$
$T_i = average\ (\hat{y}_i) - standard\ deviation\ (\hat{y}_i);$

(1)

where $n$ is the number of responses, $r_i$ is the relative importance of the response among all responses. $L_i$, $U_i$ and $T_i$ are the lower, upper and target values, respectively, which are desired for each response ($L_i \leq T_i \leq U_i$) as described in the literature (Derringer and Suich, 1980). In this study, the $r_i$ values selected for bio-$H_2$, $N_2$, and $CO_2$ were 5, 2, and 3, respectively, indicating their relative importance for a purified biogas. Bio-$H_2$ production should be maximized, while $N_2$ and $CO_2$ production should be minimized. Bio-$H_2$ concentration has the highest influence ($r_i = 5$) on biogas upgrading (Yousef et al., 2018). Pretreatments performed in the inoculum were intended to hinder methanogenesis. Thus, the $CH_4$ component was not considered in the desirability analysis.

## 3. Results and discussion

One of the major goals of the present study is to model the biogas compounds based on the VFAs. The VFAs and biogas compounds were measured over AD processing time. As reported in Table 1, the input factors of acetic acid, propionic acid, and butyric acid were 3–1672 mg/L, 0–282 mg/L, and 3–1984 mg/L, respectively, which were measured during 168.2 h in 35-time steps. The biogas compounds were also measured for each time step. The amounts of VFAs produced relatively wide ranges for modeling the responses to be used in the optimization process.

### *3.1. VFA analyses*

Fig. 2 shows the concentrations of the acetic acid, butyric acid, and propionic acid versus digestion time that were produced during biogas production. In all the cases except for the control, large amounts of VFAs were produced, indicating that anaerobic digestion progressed sufficiently. When VFAs accumulate during anaerobic digestion, they can provide resistance to greater and faster changes in pH, thus helping maintain optimal biological activity and anaerobic digestion stability (Hu et al., 2019).

In bio-$H_2$ production, the ratio of butyric acid to acetic acid (B/A ratio) and propionic acid concentration are indicators of hydrogen production; higher B/A ratios and lower concentrations of propionic acid reflect higher efficiency of bio-$H_2$ production (Chen et al. 2002; Han and Shin 2004). As shown in Fig. 2, all the pretreatments increased butyric acid and acetic acid successfully and decreased propionic acid drastically, indicating the pretreatments were suitable to increase bio-$H_2$ production.

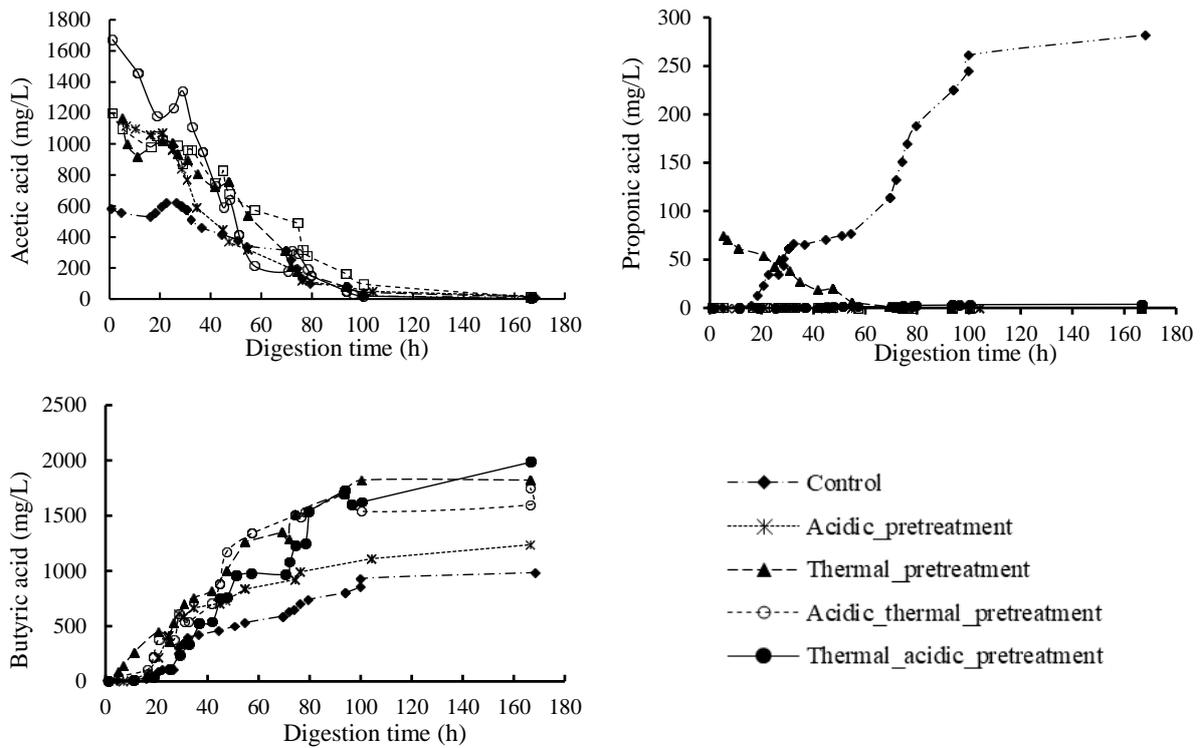

Figure 2 – Changes in VFAs versus digestion time for the pretreatments.

At lower concentrations of propionic acid and higher B/A ratios, the efficiency of hydrogen production increases. Mean amounts of propionic acid and B/A ratio of control, acidic, thermal, acidic-thermal, and thermal-acidic pretreatments were 101.93 and 7.65, 0.19 and 5.45, 22.27 and 17.31, 0.18 and 12.02, and 1.56 and 57.68, respectively. As shown, the acidic-thermal pretreatment could decrease the propionic acid, and thermal-acidic pretreatment could increase the B/A more than the other pretreatments. Therefore, these two pretreatments were better than the other pretreatments at improving the VFAs for bio-$H_2$ production. In the control, propionic acid was higher than in the other treatments, and the B/A ratio was low; therefore, the efficiency of bio-$H_2$ production was low. Although these results could explain the overall effect of VFAs on the biogas compounds, they cannot go into great detail and discover the effect of acetic acid, butyric acid, and propionic acid on biogas compounds completely. So, the biogas compounds were analyzed based on VFAs using regression models. As shown in Fig. 2, the relations between VFAs and digestion time were not linear for the assays. Similar trends existed for the biogas compounds and VFAs and digestion time.

Therefore, it is not suitable to model the biogas compounds based on the VFAs using time-independent nonlinear models. Furthermore, in the modeling process, the time cannot be used within the VFAs variables and should be used during the variables. Therefore, the bellow assumptions were done to model and optimize the responses:

- VFAs were related to the digestion time directly and nonlinearly, $x(t)$.
- Biogas compounds were related to both the VFAs and digestion time, $y(x,t)$.

### 3.2. Nonlinear regression modeling

A quadratic model (Eq. 2) was generalized for all the responses according to the VFAs.

$$y(x) = a_0 + \sum_{i=0}^{n} a_i x_i + \sum_{i=0}^{n} a_{ii} x_i^2 + \sum_{i<j}^{n} a_{ij} x_i x_j + ....x_i \,(i=1,2,....n) \tag{2}$$

where $y$ is the responses, $x_1$, $x_2$, and $x_3$ show the amounts of acetic acid, propionic acid, and butyric acid, respectively, $a_i$ is the model coefficients, and $N$ is the number of input factors ($N$=3) described in Table 1. Eq. 2 was developed using the dataset, and the final models of bio-$H_2$, $N_2$, $CH_4$, and $CO_2$ were produced as Eqs. 3–6, respectively.

$$Bio - H_2 = -8.104 + 0.018x_1 + 0.119x_2 + 0.440x_3 - 0.0002x_1^2 - 0.00002x_2^2 - 0.0004x_3^2 - 0.000008x_1x_2 + 0.009x_1x_3 - 0.00002x_2x_3 \tag{3}$$

$$N_2 = 36.815 - 0.011x_1 - 0.553x_2 - 0.052x_3 + 0.0006x_1^2 + 0.00002x_2^2 + 0.0005x_3^2 + 0.000007x_1x_2 + 0.0004x_1x_3 + 0.00002x_2x_3 \tag{4}$$

$$CH_4 = 6.698 - 0.002x_1 + 0.256x_2 - 0.0095x_3 - 0.0002x_1^2 + 0.000003x_2^2 - 0.0001x_3^2 - 0.000002x_1x_2 - 0.0005x_1x_3 + 0.000003x_2x_3 \tag{6}$$

$$CO_2 = 3.385 + 0.003x_1 + 0.441x_2 + 0.026x_3 - 0.0004x_1^2 - 0.000004x_2^2 - 0.0002x_3^2 - 0.000003x_1x_2 - 0.001x_1x_3 + 0.000008x_2x_3 \tag{5}$$

The quantities of adjusted $R^2$, $R^2$ and MSE calculated for the bio-$H_2$, $N_2$, $CH_4$, and $CO_2$ were 0.537, 0.567 and 39.831, 0.797, 0.81 and 30.285, 0.837, 0.963 and 1.555, and 0.663, 0.684 and 29.18, respectively. Predicted values of the responses using multivariate regression models versus observed values are illustrated in Fig. 3. Overall dataset of the pretreatments was used to model the biogas compounds except for $CH_4$ production, where the control dataset was used.

The pretreatments, performed in the inoculum, were intended to hinder methanogenesis. Yet, the quantities showed good accuracy for the models, these models were developed as time-independent, which is why they were not trustable to be used in the optimization process.

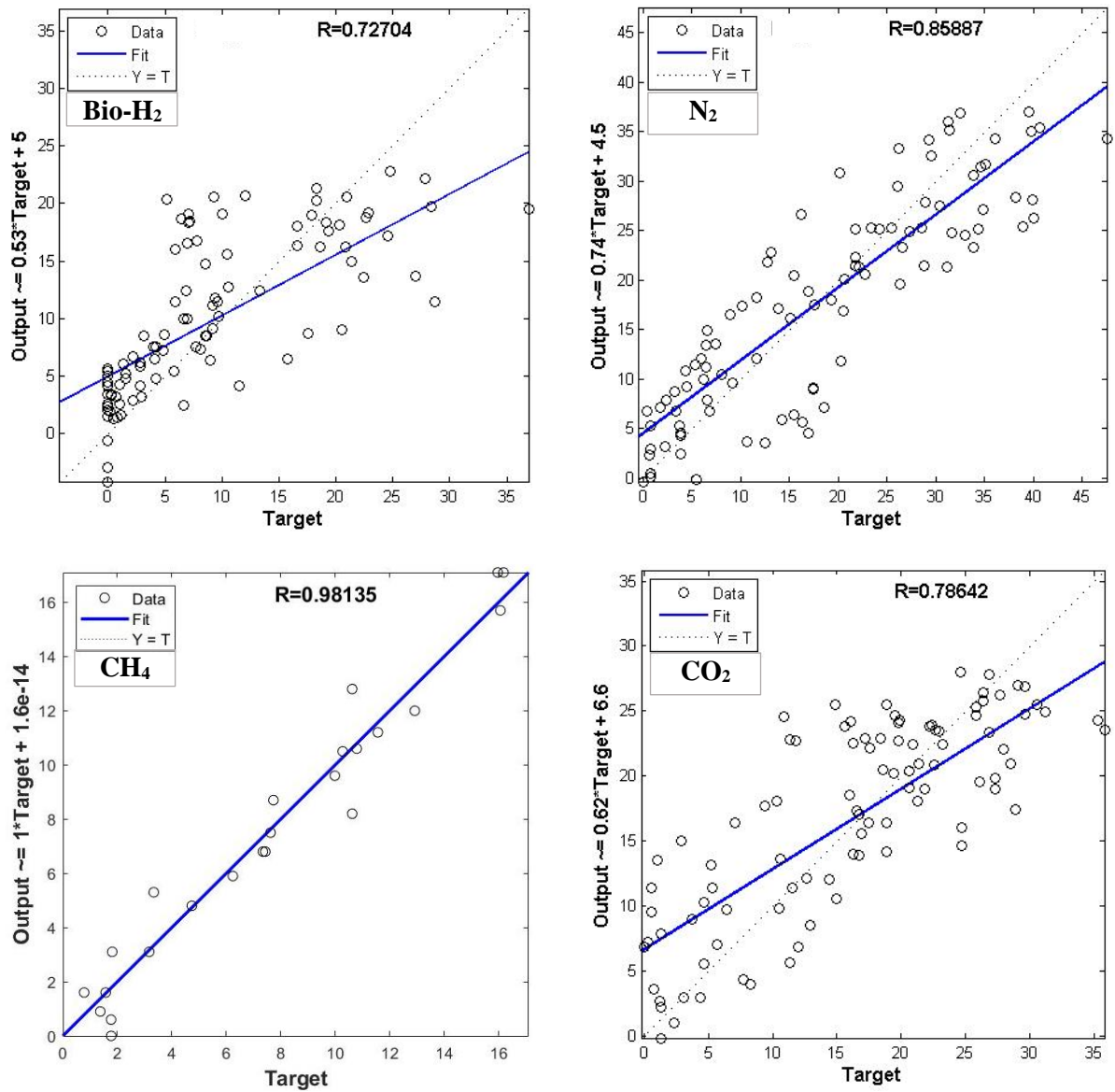

Figure 3 – Predicted amounts (output) of the responses using multivariate regression models versus observed amounts (target).

Regression analysis on the standardized dataset showed that butyric acid ($x_3$) had the greatest effect on bio-$H_2$, propionic acid ($x_2$) had the greatest effect on $CH_4$ production, and acetic acid ($x_1$) had the greatest effect on $N_2$ and $CO_2$ production. Furthermore, the analysis revealed that the effect of each VFA on the biogas compounds was higher than their interactions.

## *3.3. DNN modeling*

In the DNN model, each neuron was trained using the GA according to the VFAs of all the pretreatments for time step 1, then the trained weights and biases were updated for the next time step respectively within the specified time. Predicted amounts of the biogas compounds versus their observed amounts for the training dataset are illustrated in Fig. 4.

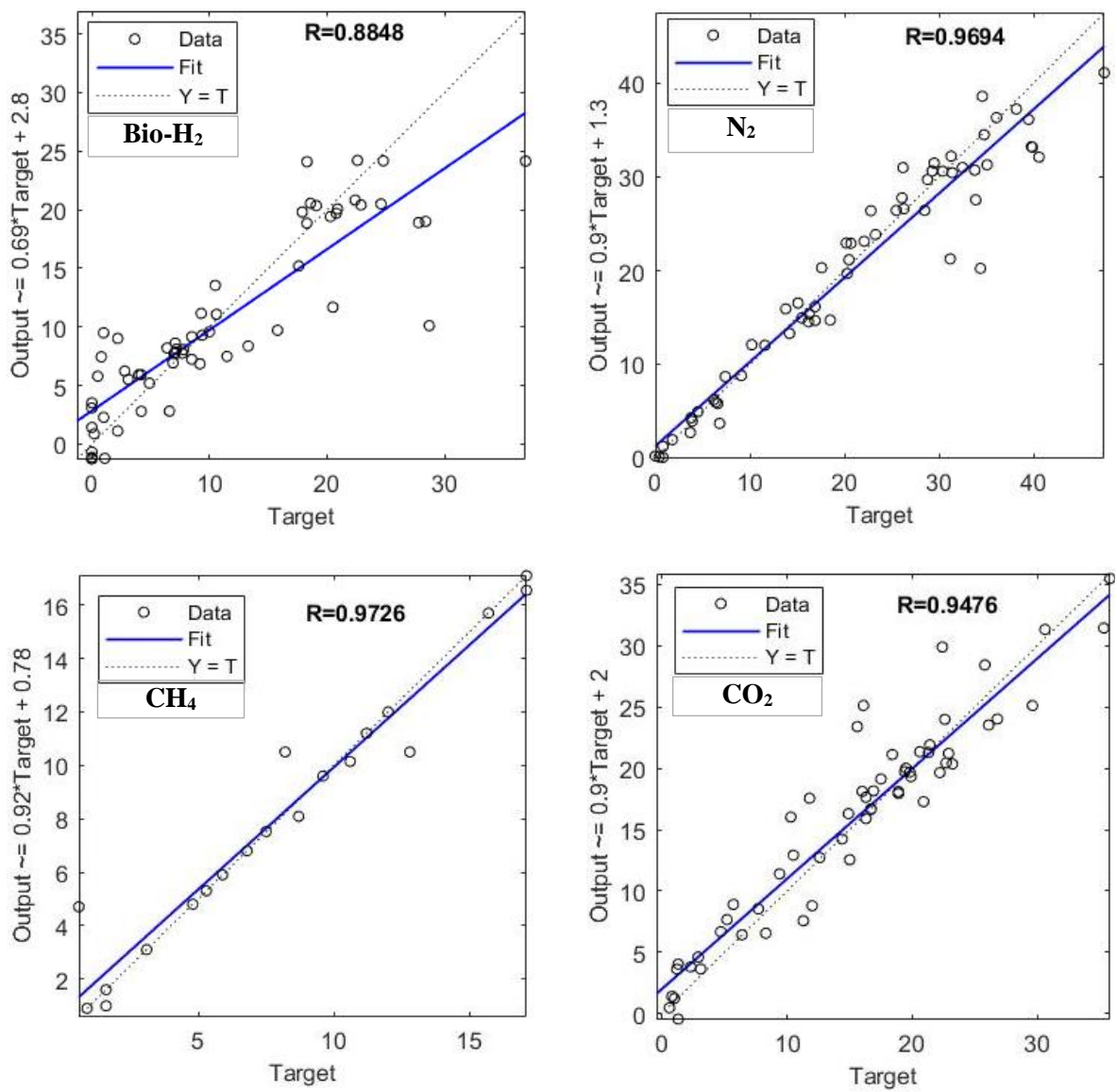

Figure 4 – Predicted amounts (output) of the responses using DNN model versus observed amounts (target).

Quantities of adjusted $R^2$, $R^2$ and MSE of the DNN model to predict the bio-$H_2$, $N_2$, $CH_4$, and $CO_2$ production for the training dataset were 0.697, 0.783 and 26.634, 0.831, 0.94 and 28.189, 0.843, 0.946 and 1.434, and 0.801, 0.898 and 23.229, respectively. Using the test dataset, the quantities of adjusted $R^2$, $R^2$ and MSE of the models of bio-$H_2$, $N_2$, $CH_4$, and $CO_2$ production were 0.701, 0.740 and 27.426, 0.798, 0.815 and 30.123, 0.934, 0.972 and 1.305, and 0.715, 0.757 and 25.364, respectively. These statistical parameters show that the accuracy of the DNN model was higher than that of the multivariate regression models. Furthermore, as described above the DNN model was retrained based on the digestion time, indicating that this model can determine the effect of time. Therefore, the predicted amounts of the DNN model were used in the desirability analysis to determine the best amounts of VFAs. According to the literature, innovative ANN approaches are highly accurate to estimate the biogas production of wastewater treatment systems (Beltramo et al., 2016; Oloko-Oba et al., 2018; Sakiewicz et al., 2020; Suberu et al., 2020; Mahmoodi-Eshkaftaki and Ebrahimi, 2021).

To determine the accuracy of the DNN model to predict the biogas compounds according to the digestion time, 10 random time steps were selected. The biogas compounds were predicted according to the VFAs amounts for these time steps. The predicted versus observed amounts of bio-$H_2$ for the time steps are illustrated in Fig. 5. As shown the DNN model was successful to predict the bio-$H_2$ over digestion time. Similar trends were found for the other biogas compounds. The results show that the MSE of the DNN model to predict the bio-$H_2$, $N_2$, $CH_4$, and $CO_2$ production according to the digestion time was 56.25, 22.37, 1.24, and 85.32, respectively.

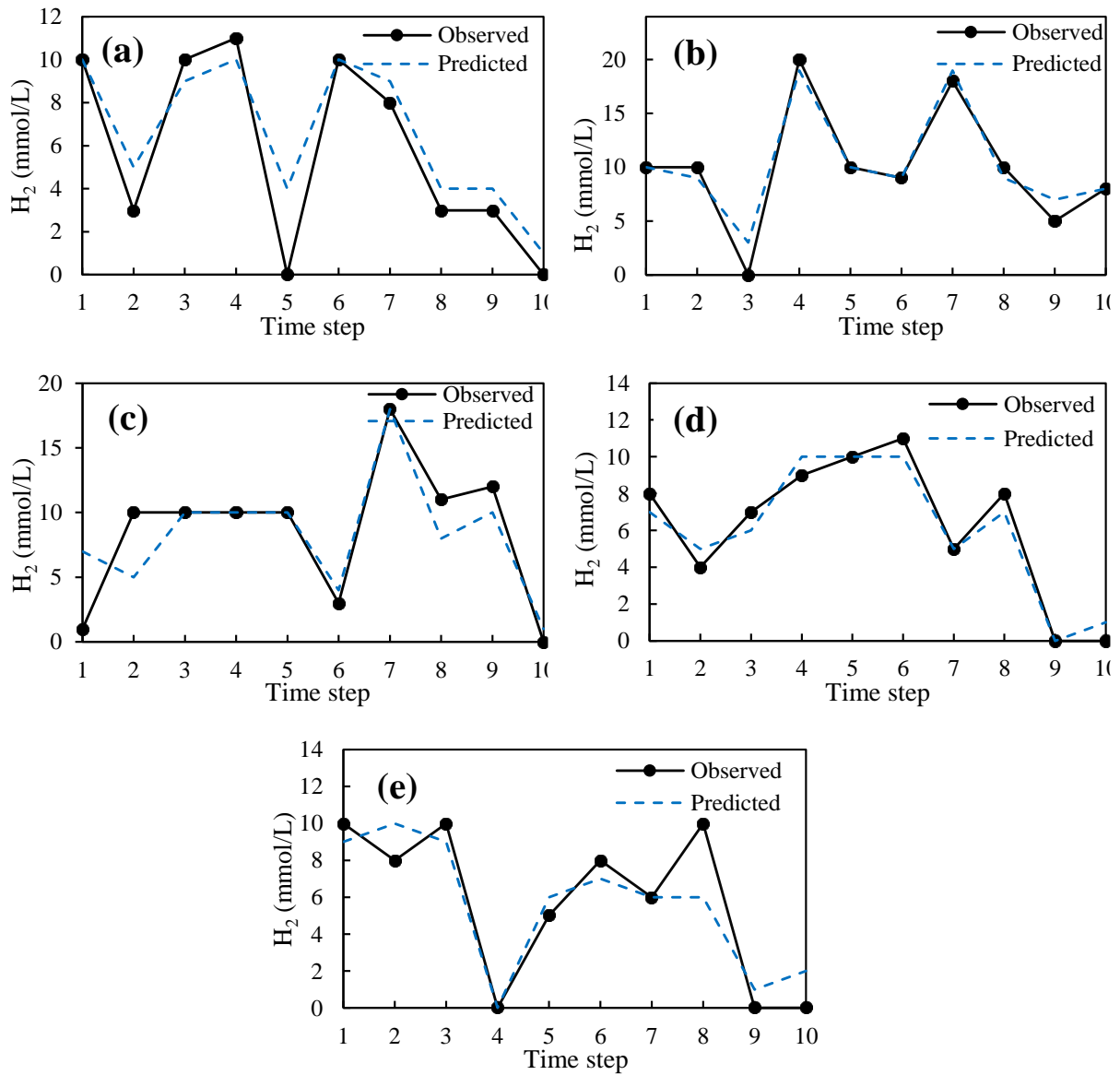

Figure 5 – Observed and predicted amounts of bio-$H_2$ for 10 random time steps (a) control, (b) acidic pretreatment, (c) thermal pretreatment, (d) acidic-thermal pretreatment, and (e) thermal-acidic pretreatment.

### 3.4. Optimization

Slurry can have the best digestion when its properties are in optimum conditions. As described above, the VFAs amounts influence the anaerobic digestion and biogas purification. The optimization of these parameters is essential for the successful operation of the anaerobic digestion, though it is difficult to carry out. The predicted amounts of biogas compounds using the DNN model over 200 time steps are shown in Fig. 6. To predict the biogas compounds with more details using the DNN model, the digestion time 0–168.2 h was divided in 200 time steps (each time step equal to 0.841 h). It is shown that bio-$H_2$ production increased during 20–50 time steps (16.82–42.05 h), and then the production remained unchanged. $N_2$ production increased with time step and $CO_2$ decreased with time step until 200 time steps. The desirability values calculated for each biogas compound in Fig. 6 show that the desirability of bio-$H_2$ production increased with time after time step 20 (16.82 h), the desirability of $N_2$ production decreased with time after time step 30 (25.23 h), and the desirability of $CO_2$ production increased with time after time step 30 (25.23 h) of digestion.

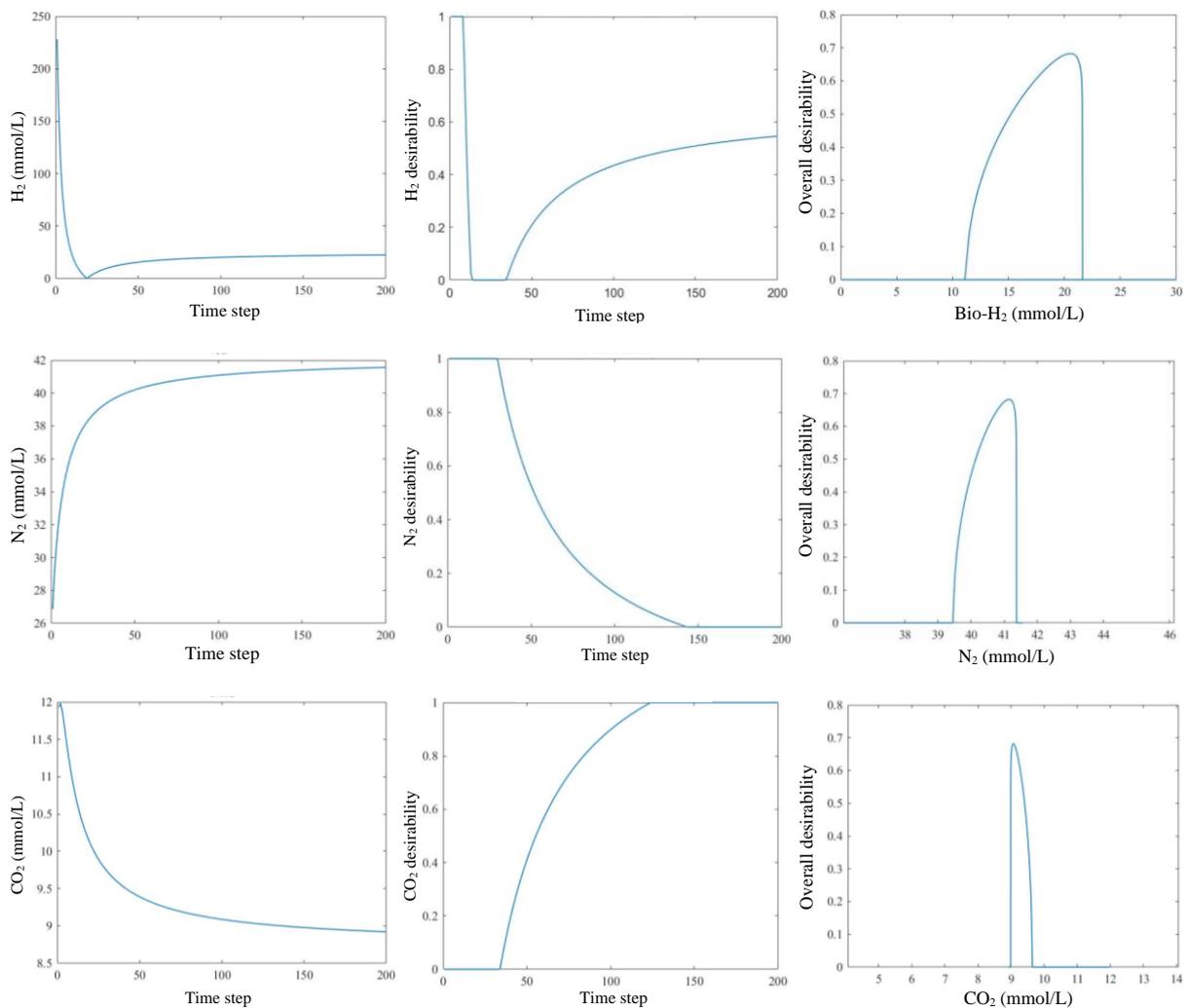

Figure 6 – Biogas compounds versus time step and their desirability values.

For desirable digestion, the biogas compounds should be optimized simultaneously using constraints. As described for the constraints, bio-$H_2$ production was maximized, while $N_2$ and $CO_2$ production were minimized. Therefore, overall desirability factor of the digestion was determined by overlaying the individual desirability values of the biogas compounds. $CH_4$ was not used in the optimization process. The overall desirability function ($D$) versus biogas compounds for the best run is illustrated in Fig. 6. The function $D$ was not quite flat in the vicinity of the optimal solution, indicating that variations around the maximum desirability could change the overall desirability drastically. The overall desirability values calculated of the predicted amounts of DNN model were maximized, indicating a good fitness for this model. As shown in this figure, the optimum ranges of bio-$H_2$, $N_2$, and $CO_2$ production were 12–22 mmol/L, 39.5–41.5 mmol/L, and 9–9.8 mmol/L, respectively, which were achieved in the digestion time 25.23–123.63 h. The optimum absolute amounts of bio-$H_2$, $N_2$, and $CO_2$ production were 20.524 mmol/L, 41.149 mmol/L, and 9.067 mmol/L, respectively.

Different runs were determined by running the DNN model in order of high desirability, from which the first six runs are reported in Table 2. These runs were used to illustrate the effect of changes in the digestion time and VFAs on the biogas compounds. The best run (Run 1) revealed the highest desirability and was selected to report the optimum absolute amounts in this research. These optimum runs were precise and reliable, because they were determined by a hybrid optimization technique, integrating an accurate model (DNN model) and desirability analysis. Similar optimization methods were developed in other studies by integrating precise regression models and desirability analysis, and their optimum amounts were accurate for practical use (Arun et al., 2017; Mahmoodi-Eshkaftaki and Rahmanian-Koushkaki, 2020; Mahmoodi-Eshkaftaki and Mockaitis, 2022).

Table 2 – Optimum absolute amounts of the digestion time, VFAs and biogas compounds, and optimum ranges determined by overlaying the optimal values of responses.

| Parameters | Run 1 | Run 2 | Run 3 | Run 4 | Run 5 | Run 6 | Optimum range |
|---|---|---|---|---|---|---|---|
| Digestion time (h) | 67.15 | 57.58 | 62.14 | 45.60 | 50.18 | 62.14 | 25.23–123.63 |
| Acetic acid (mg/L) | 823.2 | 1374.0 | 1169.9 | 1534.3 | 1409.1 | 1084.8 | 823.2–1534.3 |
| Propionic acid (mg/L) | 36.6 | 37.4 | 43.6 | 47.4 | 44.3 | 36.3 | 36.3–47.4 |
| Butyric acid (mg/L) | 1822 | 1623 | 1822 | 1522 | 1584 | 1749 | 1522–1822 |
| Bio-$H_2$ (mmol/L) | 20.52 | 13.15 | 6.41 | 15.15 | 12.80 | 5.31 | 6.41–26.25 |
| $N_2$ (mmol/L) | 41.15 | 17.84 | 27.61 | 13.60 | 12.22 | 30.82 | 12.22–43.24 |
| $CO_2$ (mmol/L) | 9.07 | 10.98 | 5.05 | 24.10 | 11.63 | 23.02 | 5.05–25.30 |
| $CH_4$ (mmol/L) | 0 | 0 | 0.33 | 0 | 0.11 | 1.37 | 0–1.40 |
| Max desirability | 0.68 | 0.65 | 0.63 | 0.58 | 0.57 | 0.55 | 0.51–0.68 |

According to Table 2, the optimum absolute amounts of acetic acid, propionic acid, and butyric acid were 823.2 mg/L, 36.6 mg/L, and 1822 mg/L, respectively, related to digestion time of 67.15 h. These VFAs resulted in an increase in bio-$H_2$ up to 20.52 mmol/L, and a decrease in $N_2$ and $CO_2$ to 41.15 mmol/L and 9.07 mmol/L, respectively. As shown, the optimum amount of $CH_4$ was 0 mmol/L, which agreed with the pretreatment analysis. As described above, the pretreatments were performed in inoculum to hinder methanogenesis. The suitable ranges of the VFAs in each run were overlayed and the optimum ranges of acetic acid, propionic acid, and butyric acid were 823.2–1534.3 mg/L, 36.3–47.4 mg/L, and 1522–1822 mg/L, respectively. These values were determined for an optimum range of digestion time 25.23–123.63 h. As found, the optimal ranges of bio-$H_2$, $N_2$, $CO_2$, and $CH_4$ were 6.4–26.2 mmol/L, 12.2–43.2 mmol/L, 5–25.3 mmol/L, 0–1.4 mmol/L, respectively. In recent years, researchers have studied the optimization of VFAs and digestion time and have reported some of the best values consistent with our research results. As reported, a maximum bio-$H_2$ production is achieved with digestion time of 30 h (Radjaram and Saravanane, 2011), 72 h (Badiei et al., 2011), and 24 h (Prabakar et al., 2018). Lin et al. (2012) found a suitable range of 0.5–72 h and Kim et al. (2013) achieved a suitable range of 16–18 h for hydraulic retention time. As reported, the slurry properties such as substrate concentration, pH, temperature, etc. affect the digestion time. Kim et al. (2013) reported that the production of hydrogen began after 3 h, peaking at 16 h of operation and declining after 18 h. They found that starting from 3–16 h of the digestion time, the ratio B/A increased, while propionic acid decreased, and as the production rate of propionic acid increased, the metabolism shifted from the production of hydrogen toward its consumption. In the literature, there is no report on the optimal amounts of VFAs to increase bio-$H_2$, only the trends of VFAs have been considered, though their optimal amounts are very important. However, in this research we studied both the trends of VFAs and their optimal values to increase bio-$H_2$ and decrease $N_2$ and $CO_2$ components. Our findings showed that both acetic acid and butyric acid increased together, while propionic acid was different. In this study, as the B/A ratio increased, propionic acid decreased; and bio-$H_2$ increased which was consistent with previous studies. As shown in Table 2, the best amounts of acetic acid and butyric acid were 823.2 mg/L and 1822 mg/L, producing the highest B/A (2.21) among the runs. Furthermore, the best propionic acid was 36.6 mg/L, producing a low amount for this acid. Our findings not only illustrate the trends of VFAs to increase the digestion efficiency but also reveal the optimum amounts of VFAs, while all the digestion components are time-dependent.

This decrease in B/A is consistent with the finding that acetate is probably the major route to produce bio-$H_2$ through the formation and subsequent degradation of ethanol, as depicted by Mockaitis et al. (2020), showing that acetate is a key compound for hydrogen production. Moreover, propionate acidogenesis consumes bio-$H_2$, thus low amounts of propionate could be usually associated with bio-$H_2$ production through dark fermentation (Peixoto et al., 2012; Mockaitis et al., 2020; Mockaitis et al., 2022), reinforcing the conclusion found in this study.

Time dependency was also an important factor for bio-$H_2$ production through dark fermentation of xylose. This dependency is corroborated through the comparison of the microbiology profile between the control assay and all experiments with a pretreated inoculum. Time dependency shows that although the pretreatments were able to halt methanogenesis in all pretreatments performed, it takes time to reestablish a new microbial community based on the selection imposed by the pretreatments. As reported by Mockaitis et al. (2020), the development of bacteria from the genus *Truepera*, and the family *Peptostreptococcaceae* was important for a more efficient bio-$H_2$ production. In this sense, it is important to highlight that all pretreatments reduced the microbial diversity, and the time dependency is most probable to be related to the growth of the selected microorganisms and the establishment of a structured microbial consortium able to degrade xylose into bio-$H_2$.

Model showed that the optimum response for bio-$H_2$ production occurred in time range between 6.41-26.25 h, which coincides with the experimental finding for the exponential growth phase in acidic pretreatment, which was the best inoculum pretreatment for bio-$H_2$ production. Mockaitis et al. (2020) found similar values modeling the bio-$H_2$ production using a growth and kinetic-based model.

In this study we managed the appropriate range of the VFAs to improve the efficiency of biogas plant over digestion time despite difficulties. It is practically impossible to keep the VFAs at constant levels to improve anaerobic digestion. Even though it is possible to control the environment, by using equipment to a certain extent, this would considerably increase the cost of operations. In this sense, the design of smaller bioreactors operating at a high rate and efficiency is better than usual anaerobic ponds or lagoons, since bioreactors are easier to control and more cost effective. Range optimization by integrating the neural network and desirability analysis techniques based on overlaying the parameters is a convenient tool for the operators in such situations. These facts point to the importance of neural network modeling, desirability function and overlaying method in this study.

## 4. Conclusion

In this study, the relationship between VFAs and biogas compounds was studied to determine the optimum ranges of the VFAs during the digestion time. All the parameters were studied as time-dependent components in five different pretreatments. In all the cases except for the control, large amounts of VFAs were produced, indicating that anaerobic digestion progressed sufficiently. All the pretreatments in the inoculum successfully increased butyric and acetic acids production and decreased propionic acid production drastically, indicating that the pretreatments were completely suitable to increase the efficiency of bio-$H_2$ production. Acidic-thermal pretreatment decreased propionic acid and thermal-acidic pretreatment increased B/A ratio more than the other pretreatments, confirming that these two pretreatments were better able to improve the VFAs for bio-$H_2$ production than the other pretreatments. Multivariate regression analysis showed that butyric acid had the greatest effect on bio-$H_2$, and propionic acid had the greatest effect on $CH_4$ production, while the acetic acid had the greatest effect on $N_2$ and $CO_2$ production. The development of an accurate DNN model results in higher accuracy than the multiple regression model. In addition, it can navigate in time and space, which can improve the optimization process of bio-$H_2$ production. Through training and testing the DNN model, the accuracy of the DNN model was found to be very high, indicating that the model is suitable for the optimization process. Therefore, the DNN model was integrated with desirability analysis to determine the optimum amounts of the VFAs and biogas compounds. Our results show that as the B/A ratio increases and propionic acid decreases, bio-$H_2$ increases. The highest efficiency of biogas production or bio-$H_2$ production was achieved with the highest B/A (2.21) determined for acetic acid 823.2 mg/L and butyric acid 1822 mg/L, and a low amount of propionic acid 36.6 mg/L. The highest efficiency was achieved for digestion time of 67.15 h, producing the highest amount of bio-$H_2$ (20.52 mmol/L) and a low amount of $N_2$ (41.15 mmol/L) and $CO_2$ (9.07 mmol/L). This study used the integrated DNN model and desirability function to optimize the VFAs based on digestion time and upgrade biogas purification. Additionally, it is possible to infer that time-dependent models are more able to describe the system since these models consider microbial growth and community selection. It is recommended to use the pareto genetic algorithm instead of the desirability function to optimize all responses at the same time, and to increase the generalization of the optimal value in future research.

**References**

Arun, V.V., Saharan, N., Ramasubramanian, V., Babitha Rani, A.M., Salin, K.R., Sontakke, R., Haridas, H., Pazhayamadom, D.G., 2017. Multi-response optimization of Artemia hatching process using split-split-plot design-based response surface methodology. Sci. Rep. 7, 1–13. https://doi.org/10.1038/srep40394,40394.

Badiei, M., Jahim, J.M., Anuar, N., Sheikh Abdullah, S.R., 2011. Effect of hydraulic retention time on biohydrogen production from palm oil mill effluent in anaerobic sequencing batch reactor. Int. J. Hydrog. Energy 36, 5912–5919. https://doi.org/10.1016/j.ijhydene.2011.02.054.

Beltramo, T., Ranzan, C., Hinrichs, J., Hitzmann, B., 2016. Artificial neural network prediction of the biogas flow rate optimized with an ant colony algorithm. Biosyst. Eng. 143, 68–78. https://doi.org/10.1016/j.biosystemseng.2016.01.006.

Chen, C.C., Lin, C.Y., Lin, M.C., 2002. Acid-base enrichment enhances anaerobic hydrogen production process. Appl. Microbiol. Biotechnol. 58, 224–228. https://doi.org/10.1007/s002530100814.

Derringer, G., Suich, R., 1980. Simultaneous optimization of several response variables. J. Qual. Technol. 12, 214–219. https://doi.org/10.1080/00224065.1980.11980968.

Fan, M., Hu, J., Cao, R., Xiong, K., Wei, X., 2017. Modeling and prediction of copper removal from aqueous solutions by nZVI/rGO magnetic nanocomposites using ANN-GA and ANN-PSO. Sci. Rep. 7, 18040. doi: 10.1038/s41598-017-18223-y.

Feng, S., Zhou, H., Dong, H., 2019. Using deep neural network with small dataset to predict material defects. Mater. Des. 162, 300–310. https://doi.org/10.1016/j.matdes.2018.11.060.

Goodfellow, I., Bengio, Y., Courville, A., 2016. Deep Learning. MIT press, Cambridge.

Han, S.K., Shin, H.S., 2004. Biohydrogen production by anaerobic fermentation of food waste. Int. J. Hydrog. Energy 29, 569–577. https://doi.org/10.1016/j.ijhydene.2003.09.001.

Hu, Y., Wu, J., Li, H., Poncin, S., Wang, K., Zuo, J., 2019. Novel insight into high solid anaerobic digestion of swine manure after thermal treatment: kinetics and microbial community properties. J. Environ. Manag. 235, 169–177. doi: 10.1016/j.jenvman.2019.01.047

Izumi, K., Okishio, Y.K., Nagao, N., Niwa, C., Yamamoto, S., Toda, T., 2010. Effects of particle size on anaerobic digestion of food waste. Int. Biodeterior. Biodegrad. 64 (7), 601–608.

Kegl, T., Kralj, A.K., 2020. Multi-objective optimization of anaerobic digestion process using a gradient-based algorithm. Energy Convers. Manag. 226, 113560.

Kim, S., Choi, K., Kim, J.O., Chung, J., 2013. Biological hydrogen production by anaerobic digestion of food waste and sewage sludge treated using various pretreatment technologies. Biodegradation 24, 753–764. doi: 10.1007/s10532-013-9623-8.

Liu, B., Wei, Y., Zhang, Y., Yang, Q., 2017. Deep neural networks for high dimension, low sample size data. Proceedings of the Twenty-Sixth International Joint Conference on Artificial Intelligence (IJCAI-17), Melbourne, Australia, pp. 65−70.

Lin, Y.H., Zheng, H.X., Juan, M.L., 2012. Biohydrogen production using waste activated sludge as a substrate from fructose-processing wastewater treatment. Process Saf. Environ. Prot. 90, 221–230. https://doi.org/10.1016/j.psep.2012.02.004.

Mahmoodi-Eshkaftaki, M., Ebrahimi, R., 2019. Assess a new strategy and develop a new mixer to improve anaerobic microbial activities and clean biogas production. J. Clean. Prod. 206, 797–807. https://doi.org/10.1016/j.jclepro.2018.09.024.

Mahmoodi-Eshkaftaki, M., Ebrahimi, R., 2021. Integrated deep learning neural network and desirability analysis in biogas plants: A powerful tool to optimize biogas purification. Energy 231, 121073. https://doi.org/10.1016/j.energy.2021.121073.

Mahmoodi-Eshkaftaki, M., Mockaitis, G., 2022. Structural optimization of biohydrogen production: impact of pretreatments on volatile fatty acids and biogas parameters. Int. J. Hydrog. Energy. 47, 7072–7081. https://doi.org/10.1016/j.ijhydene.2021.12.088.

Mahmoodi-Eshkaftaki, M., Rahmanian-Koushkaki, H., 2020. An optimum strategy for substrate mixture and pretreatment in biogas plants: potential application for high-pH waste management. Waste Manag. 113, 329–341. https://doi.org/10.1016/j.wasman.2020.06.014.

Mockaitis, G., Bruant, G., Guiot, S.R., Peixoto, G., Foresti, E., Zaiat, M., 2020. Acidic and thermal pretreatments for anaerobic digestion inoculum to improve hydrogen and volatile fatty acid production using xylose as the substrate. Renew. Energy 145, 1388–1398. https://doi.org/10.1016/j.renene.2019.06.134.

Mockaitis, G., Bruant, G., Foresti, E., Zaiat, M., Guiot, S.R., 2022. Physicochemical pretreatment selects microbial communities to produce alcohols through metabolism of volatile fatty acids. Biomass Conversion and Biorefinery. https://doi.org/10.1007/s13399-022-02383-7.

Oloko-Oba, M.I., Taiwo, A.E., Ajala, S.O., Solomon, B.O., Betiku, E., 2018. Performance evaluation of three different-shaped bio-digesters for biogas production and optimization by artificial neural network integrated with genetic algorithm. Sustain. Energy Technol. Assessments 26, 116–124. https://doi.org/10.1016/j.seta.2017.10.006.

Paudel, S.R., Banjara, S.P., Choi, O.K., Park, K.Y., Kim, Y.M., Lee, J.W., 2017. Pretreatment of agricultural biomass for anaerobic digestion: Current state and challenges. Bioresour. Technol. 245, 1194–1205. http://dx.doi.org/10.1016/j.biortech.2017.08.182.

Peixoto, G., Pantoja-Filho, J.L.R., Agnelli, J.A.B., Barboza, M., Zaiat, M., 2012. Hydrogen and methane production, energy recovery, and organic matter removal from effluents in a two-stage fermentative process. Appl. Biochem. Biotechnol., 168, 651-671. http://dx.doi.org/10.1007/s12010-012-9807-4.

Radjaram, B., Saravanane, R., 2011. Assessment of optimum dilution ratio for biohydrogen production by anaerobic co-digestion of press mud with sewage and water. Bioresour. Technol. 102, 2773–2780. doi: 10.1016/j.biortech.2010.11.075.

Rambabu, K., Bharath, G., Banat, F., Hai, A., Show, P.L., Nguyen, T.H.P., 2021. Ferric oxide/date seed activated carbon nanocomposites mediated dark fermentation of date fruit wastes for enriched biohydrogen production. Int. J. Hydrog. Energy 46, 16631–16643. https://doi.org/10.1016/ j.ijhydene.2020.06.108.

Ranjan, D., Mishra, D., Hasan, S.H., 2011. Bio-adsorption of arsenic: an artificial neural networks and response surface methodological approach. Ind. Eng. Chem. Res. 50, 9852–9863. doi: dx.doi.org/10.1021/ie200612f.

Sakiewicz, P., Piotrowski, K., Ober, J., Karwot, J., 2020. Innovative artificial neural network approach for integrated biogas – wastewater treatment system modeling: effect of plant operating parameters on process intensification. Renew. Sust. Energ. Rev. 124, 109784. https://doi.org/10.1016/j.rser.2020.109784.

Steinbusch, K.J.J., Hamelers, H.V.M., Buisman, C.J.N., 2008. Alcohol production through volatile fatty acids reduction with hydrogen as electron donor by mixed cultures. Water Res. 42, 4059–4066. https://doi.org/10.1016/j.watres.2008.05.032.

Suberu, C.E., Kareem, K.Y., Adeniran, K.A., 2020. Artificial neural network modeling of biogas yield from co-digestion of poultry droppings and cattle dung. KUSET 14, 1–6.

Taherzadeh, M.J., Karimi, K., 2008. Pretreatment of lignocellulosic wastes to improve ethanol and biogas production: a review. Int. J. Mol. Sci. 9, 1621–1651. doi: 10.3390/ijms9091621.

Thanwised, P., Wirojanagud, W., Reungsang, A., 2012. Effect of hydraulic retention time on hydrogen production and chemical oxygen demand removal from tapioca waste- water using anaerobic mixed cultures in anaerobic baffled reactor (ABR). Int. J. Hydrog. Energy 37, 15503–15510. https://doi.org/10.1016/j.ijhydene.2012.02.068.

Wang, H., Li, J., Cheng, M., Zhang, F., Wang, X., Fan, J., Wu, L., Fang, D., Zou, H., Xiang, Y., 2019. Optimal drip fertigation management improves yield, quality, water and nitrogen use efficiency of greenhouse cucumber. Sci. Hortic. 243, 357–366. https://doi.org/10.1016/j.scienta.2018.08.050.

Ward, L., Agrawal, A., Choudhary, A., Wolverton, C., 2016. A general-purpose machine learning framework for predicting properties of inorganic materials. NPJ Comput. Mater. 2, 16028. doi: 10.1038/npjcompumats.2016.28.

Yousef, A.M., El-Maghlany, W.M., Eldrainy, Y.A., Attia, A., 2018. New approach for biogas purification using cryogenic separation and distillation process for $CO_2$ capture. Energy 156, 328–351. https://doi.org/10.1016/j.energy.2018.05.106.